\documentclass[aps,prl,reprint,superscriptaddress]{revtex4-2}
\usepackage{amssymb}
\usepackage{amsmath,bm}
\usepackage{graphicx,color}
\usepackage[dvipsnames]{xcolor}
\usepackage{bbm}
\usepackage[bottom]{footmisc}
\usepackage{multirow}
\usepackage{soul}

\usepackage[normalem]{ulem}
\usepackage{hyperref}
\hypersetup{
    pdftoolbar=true,
    pdfmenubar=true,
    bookmarksopen=true,
    bookmarksnumbered=true,
    breaklinks=true,
    linktocpage,
    colorlinks=true,
    linkcolor={blue!95!black},
    citecolor={teal!70!black},
    urlcolor={blue!95!black},
}

\begin{document}
\title{Finite Disorder Critical Point in the Brittle-to-Ductile Transition of Amorphous Solids in the Presence of Particle Pinning}

\author{Anoop Mutneja$^\#$}
\email{anoop.mutneja2011@gmail.com} 
\affiliation{Department of Materials Science and Engineering, University of Illinois, Urbana, IL 61801, USA}
\affiliation{Materials Research Laboratory, University of Illinois, Urbana, IL, 61801, USA}
\author{Bhanu Prasad Bhowmik$^\#$}
\email{bhowmikbhanuprasad592@gmail.com}
\thanks{\\$^\#$Equal contributions} 
\affiliation{
School of Engineering, The University of Edinburgh,
King’s Buildings, Edinburgh EH9 3FG, United Kingdom}
\author{Smarajit Karmakar}
\email{smarajit@tifrh.res.in}
\affiliation{
Tata Institute of Fundamental Research, 36/P, Gopanpally Village, Serilingampally Mandal,Ranga Reddy District,
Hyderabad, Telangana 500107, India}

\begin{abstract}{The mechanical yielding of amorphous solids under external loading can be broadly classified into ductile and brittle types, depending on whether their macroscopic stress response is smooth or abrupt, respectively. Recently, it has been shown that these two regimes, obtained by tuning the degree of annealing of the system, are separated by a critical point at a finite  inherent disorder strength. Here, we demonstrate a transition from brittle yielding to ductile yielding by introducing quenched disorder in the form of randomly pinned particles. The well-annealed samples, which exhibit brittle yielding, undergo a transition to increasingly ductile  yielding with increasing pinning concentrations while exhibiting an enhanced stress overshoot. Extensive finite size analysis is performed to demonstrate the critical nature of the transition at a finite pinning concentration and the various scaling exponents obtained are found to be in good agreement with the reported values. 
Finally, we show a direct correspondence with the shear band width and the critical pining concentration to establish a possible connection between inherent disorder strength and quenched disorder strength due to particle pinning.}
   
\end{abstract}

\maketitle

\noindent{\bf Introduction:}
The mechanical failure of amorphous solids due to external deformation or applied stress is frequently observed in daily life and various natural phenomena, ranging from the breaking of car wind-shields and the flow of toothpaste from a tube to sudden landslides~\cite{RMP2018,YieldStressRMP}. In a strain-controlled scenario, on the one hand, when the applied strain is small, stress increases linearly but is often punctured by stress drops due to irreversible (plastic) rearrangements of the constituent particles~\cite{KLP_PRE2010,KLP&GeorgePRE2011}. On the other hand, at large strains, a proliferation of such plastic events occurs, leading to the saturation of stress as strain increases; a phenomenon commonly known as the yielding transition. Based on the nature of the macroscopic response, the yielding transition can be divided into two distinct classes: ductile and brittle yielding. In ductile yielding, material flow occurs gradually via plastic events distributed relatively homogeneously throughout the system, as seen in the flow of various yield-stress materials, such as foams~\cite{foamPRLLauridsen}. In contrast, the brittle yielding involves catastrophic failure through the sudden formation of a system-spanning shear band instability~\cite{RatulShearBand}, where plasticity remains concentrated in specific regions. A typical example of such failure is the breaking of metallic glass with shear band formation~\cite{GREER201371}. 

In recent years, the ductile and brittle failure of amorphous solids have become a focus of intense research~\cite{DBT_Metallic,Luo2016,D3SM01740K,DKPZ_PRE_2011,Peterlik2006,
DuctileToBrittlePNAS,OzawaPRR2022,Fielding2020DtoB,DuctileAndBrittleParley,
HimanshuSrikanthPNAS,EdanCoradoDToB,LangerPRE2020,LangerPRE2020}. In Ref.~\cite{DuctileToBrittlePNAS}, using computer simulations of glass-forming model systems, under athermal quasi-static straining (AQS) scenario~\cite{MaloneyAndLemaitre}, it has been demonstrated that the same materials can fail in either manner depending on their mechanical stability. Samples prepared with a high cooling rate or equilibrated at a high temperature fail in a ductile manner, showing a continuous stress vs. strain curve in the thermodynamic limit. In contrast, samples prepared with a low cooling rate or equilibrated at a lower temperature exhibit brittle failure, characterized by a discontinuous variation of stress  at yield point . These two types of failure are controlled by the disorder strength of the samples and are separated by a critical point at a finite disorder strength  in the context of the random field Ising model \cite{RossiRandomField}. However, this finding is challenged in Refs.~\cite{Fielding2020DtoB,Fielding2022DToB, EdanCoradoDToB}, which propose  that in the thermodynamic limit, under quasi static driving, any athermal amorphous solid with any amount of stress-overshoot exhibits a discontinuous stress drop at yielding, thus always failing in a brittle manner.  This suggests that observation of such critical point is a system size effect. However, in Ref.~\cite{RossiEPMDToB}, using an elastoplastic model~\cite{EPM1,EPM2}, that allows to simulate a very large system, the existence of this critical point is substantiated. Furthermore, the critical exponents obtained in this study are compared with those determined for the continuous-to-discontinuous transition in the AQS driven Random Field Ising Model (RFIM) with anisotropic Eshelby~\cite{Eshelby1957} type interactions~\cite{RossiPRB}, suggesting that both transitions might belong to the same universality class. Nevertheless, the ductile-to-brittle transition can be obtained by adding different types of impurities~\cite{BCK_PRL_2019,AnoopRod,ShangPengPinningPRB,RishabhSmarajitNaturePhysics} or through cyclic deformation~\cite{HimanshuSrikanthPNAS,CyclicShearBerthier,LiuMartensJCP2022}. Therefore, one can ask whether such a ductile-to-brittle transition is also separated by a critical point, and if so, whether it  belongs to the random field Ising model universality class?

In this Letter, we study brittle-to-ductile transition in amorphous solids in the presence of randomly pinned particles. The effect of particle pinning on the mechanical response of amorphous solids has been studied recently~\cite{BCK_PRL_2019,itamarPinning,ShangPengPinningPRB}. It was found that due to particle pinning, the yielding transition becomes more homogeneous, and the critical nature of the transition~\cite{JaiswalPRL2016, ItamarPNAS} is altered~\cite{BKPR_PRE_2019}.
Moreover, in the presence of pinning, the failure can be ductile even with a large stress overshoot, in contrast to what is claimed in Ref.~\cite{Fielding2020DtoB}. 

Here, we demonstrate that, starting from highly annealed samples that fail in a brittle manner, with particle pinning, the failure process loses its brittle nature.  Eventually, at a large enough pinning concentration, the failure becomes ductile. We try to identify the existence of any critical point that separates these two types of failure. Our study shows that in the presence of particle pinning, the brittle-to-ductile transition is critical. Using finite size scaling, the critical exponents are obtained. We also demonstrate a possible connection between the pinned particles as an additional disorder and the inherent annealing-dependent disorder, as well as between the shear band width and the critical length scale corresponds the mean inter-pinned particle distance.
\vskip +0.1in
\noindent{\bf Simulation details:}
We simulate a polydisperse system~\cite{NinarelloSwapPRX} of $N$ repulsive point particles, within a cubic box of dimension  $L$, with an interacting potential between $i^{th}$ and $j^{th}$ particles given by 
\begin{equation}
V_{ij}(r_{ij}) = \epsilon \left(\frac{\sigma_{i,j}}{r_{ij}}\right)^{12 } \text{.}
\label{defphi}
\end{equation}
Here $\sigma_{ij} =\left(\sigma_i+\sigma_j\right) \left(1-0.2 |\sigma_i-\sigma_j|\right)/2$.
The non-additivity is introduced to prevent crystallization. $\sigma_i$ for the $i^{th}$ particle is chosen from the following distribution, 
\begin{equation}
P(\sigma) = \frac{A}{\sigma^3}\ , \quad \sigma_{\rm min} \le \sigma \le \sigma_{\rm max}\ , 
\label{eqn2}
\end{equation}
where $A$ is a normalization constant, $\sigma_{\rm min}=0.73$ and $\sigma_{\rm max}=1.62$. The unit of the length is $\bar{\sigma} = \int_{\sigma_{min}}^{\sigma_{max}} d\sigma P(\sigma) \sigma =$ 1 and the energy scale is $\epsilon = 1$. 

To prepare a well-annealed sample first we equilibrate the system at a low~\cite{NinarelloSwapPRX} temperature $T_e = 0.06$ using swap Monte Carlo~\cite{GrigeraAndParisiSwap,SK_IP_Swap}, and then quench it to an inherent state by minimizing the energy by conjugate gradient method. To deform the system we apply strain in $xy$ plane within the athermal quasi static limit which consists of two parts. First, an affine transformation is applied to the system following $x_i = x_i + \Delta \gamma y_i$, $y_i = y_i$ and $z_i = z_i$, with $\Delta \gamma$ the strain increment taken as $5\times 10^{-5}$. Second, to keep the system in mechanical equilibrium we minimize the energy. For system with finite pinned particle concentration $c_p$, we randomly select $Nc_p$ particles from the quenched state. The same quenched states are used for all pinning concentrations.
%
%
While straining, these pinned particles participate in affine transformation but remain frozen during relaxation via energy minimization~\cite{BCK_PRL_2019}. Moreover, to avoid any kind of overlap among the pinned particles due to straining, the distance between a pair of pinned particles in the y-direction is kept greater than their interaction range. We simulate seven different system sizes $N\in[100, 400000]$ 
with data averaged over approximately  $200,~200, ~200, ~75, ~75, ~64, ~40$ and $40$ realizations, respectively.

\begin{figure}
	\includegraphics[width=0.45\textwidth]{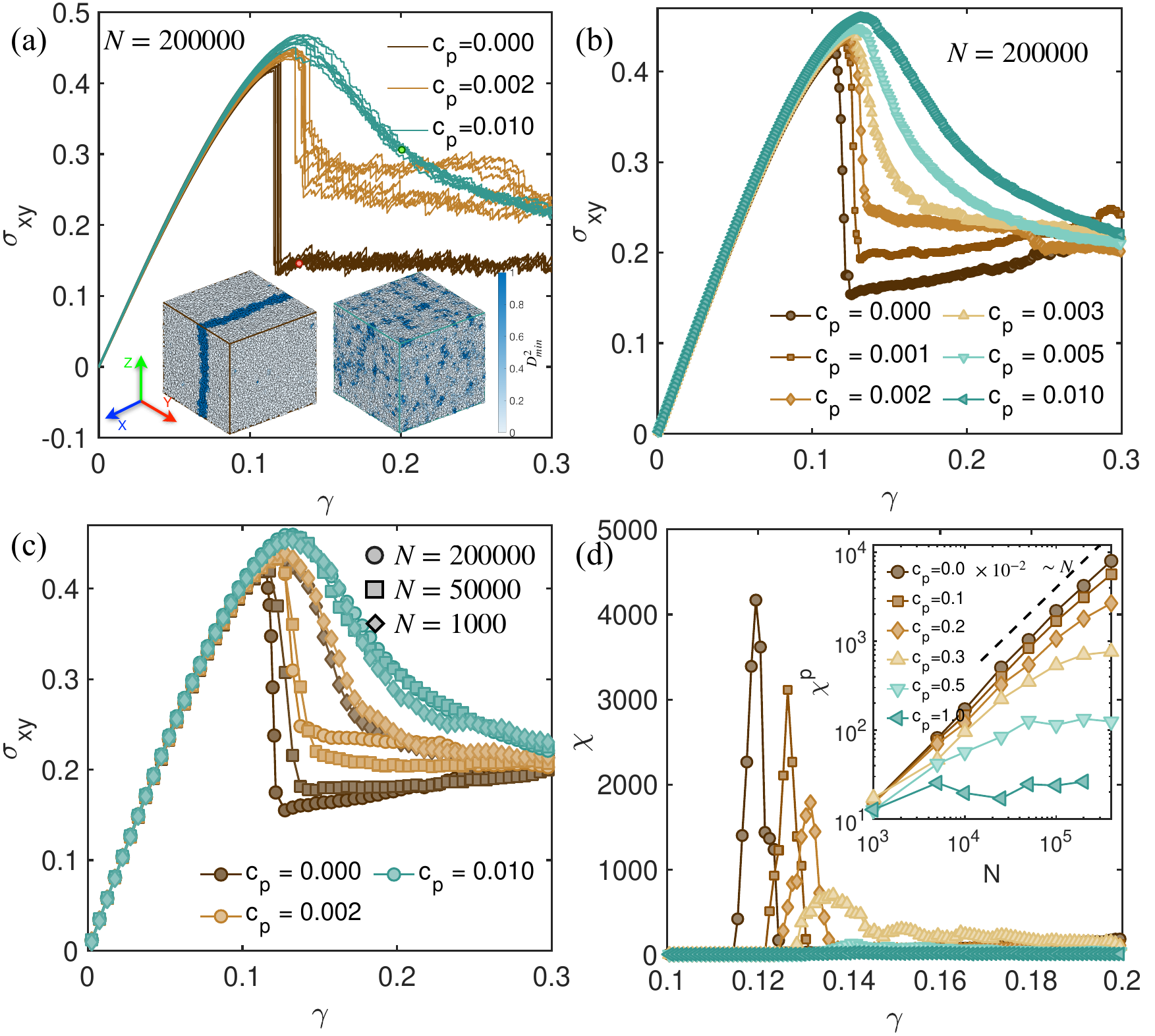}
	\caption{{\bf Brittle-to-ductile transition with increasing random pinning.} Shown are, (a) Stress $(\sigma_{xy})$ vs. strain $(\gamma)$ curves for three different pinning concentrations $c_p=0,~0.002$ and $~0.01$, each with ten different realizations. Inset: non-affine displacement map, $D^2_{min}$ \cite{Falk1998} for $c_p=0$ (left) at strain $\gamma=0.12$  and $c_p=0.01$ at $\gamma=0.2$. The unpinned system yields via formation of shear band whereas pinned system yields homogeneously. (b) $c_p$ dependence of averaged stress vs. strain curves. (c) Averaged stress vs. strain curves  for the same $c_p$ as in (a) each with three different system sizes $N$. (d) The susceptibility, $\chi=N(<\sigma_{xy}^2>-<\sigma_{xy}>^2)$, plotted as a function of strain. Inset: the peak value $\chi^{p}$ as a function of $N$.   
	} 
	\label{fig1}
\end{figure}

\begin{figure*}
	\includegraphics[width=1.0\textwidth]{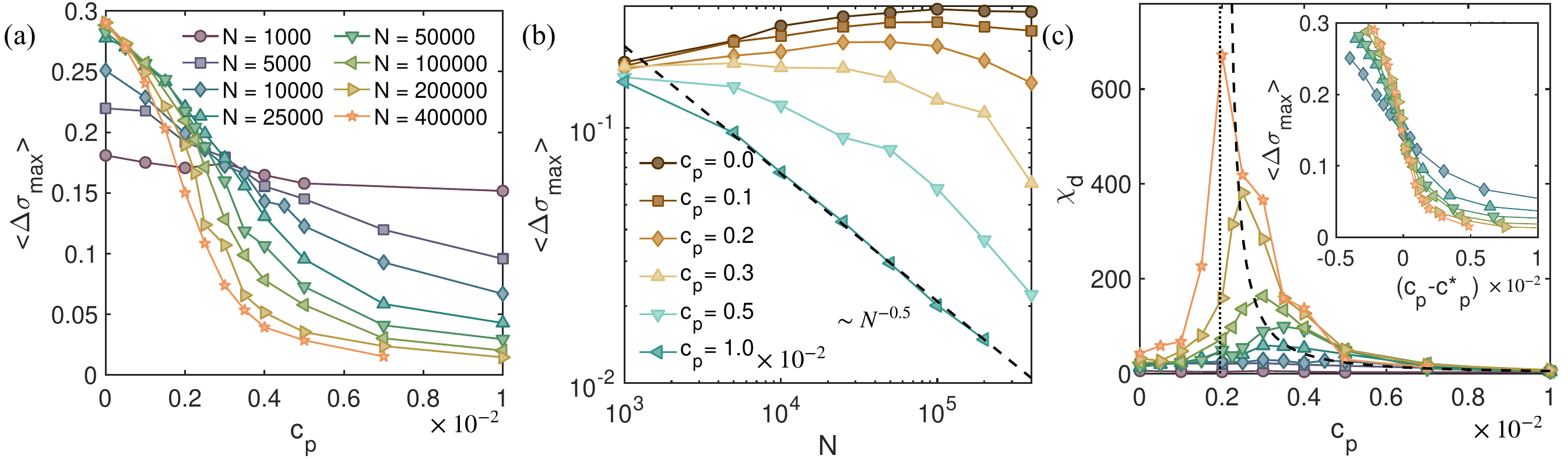}
	\caption{{\bf Order parameter for the brittle-to-ductile transition.}
(a) The order parameter, $\langle \Delta \sigma_{max}\rangle$, is plotted as a function of pinning concentration $c_p$ for different system sizes $N$. (b) The system size dependence of  $\langle \Delta \sigma_{max}\rangle$ exhibits a decay of $\langle \Delta \sigma_{max}\rangle \sim N^{-0.5}$ for $c_p=0.01$ , indicating ductile behaviour, while for unpinned or smaller $c_p$, it saturates, suggesting brittle failure in the thermodynamic limit. 
		(c) The susceptibility,  $\chi_d = N( \langle \Delta \sigma_{max}^2\rangle - \langle \Delta \sigma_{max} \rangle^2)$, as a function of  $c_p$ shows a sharp peak at $c_p^*(N)$. 
 The dashed line represents the best fit to the peak magnitude,  $\chi_d^p(N)$, vs. the peak location $c_p^*(N)$ data, using the function $\chi_d^p\sim (c_p^* - c_p^{*\infty})^{-0.73}$. The dotted line marks  $c_p^{*\infty}$. Inset shows $\langle \Delta \sigma_{max}\rangle$ as a function of rescaled pinning concentrations.  
	} 
	\label{fig2}
\end{figure*}
 \noindent{\bf Results:}
Fig.~\ref{fig1} shows the change in the macroscopic response of the system under shear deformation as a result of the incorporation of random pinning. In (a), the shear stress ($\sigma_{xy}$) vs. strain ($\gamma$) curves are shown for three distinct $c_p$, each with ten different realizations. Given that the samples are equilibrated at a very low temperature ($T_e=0.06$), the unpinned system exhibits brittle failure characterized by a large stress drop and the development of a shear band (as shown in the inset). As $c_p$ increases, the size of the stress drop reduces, and eventually, the stress-strain curves smoothen with minimal stress drops, indicative of ductile failure. 
Notably, unlike the ductile-to-brittle transitions seen with varying annealing ~\cite{DuctileToBrittlePNAS, EdanCoradoDToB,RossiEPMDToB}, here ductile failure occurs with a more pronounced stress overshoot and increased shear modulus.%

This is illustrated more clearly in the ensemble-averaged stress-strain response in Fig.~\ref{fig1}(b), which shows the effect of increasing pinning concentration in the range $c_p\in[0-0.01]$. The brittle response, marked by a discontinuous stress drop, has nearly disappeared with the introduction of pinning as low as $c_p=0.003$.
%
Fig.~\ref{fig1}(c) shows the system size dependence (marked by different symbols) of ensemble-averaged stress-strain response for the same $c_p$ as in (a) (marked by different colours). For unpinned systems, the failure becomes sharper with increasing system size, supporting the discontinuous brittle nature. For $c_p = 0.01$, we do not observe such dependence on $N$. 
The discontinuity in stress evolution is typically quantified using the susceptibility, defined as $\chi(\gamma)=\langle\sigma_{xy}^2(\gamma) \rangle-\langle \sigma_{xy}(\gamma)\rangle^2$. It is shown in Fig.~\ref{fig1}(d) for various $c_p$. For brittle systems undergoing a first-order nonequilibrium yielding transition, the peak of $\chi$ shows a linear increase with system size, as depicted in the inset for $c_p=0$. Additionally, the saturation of the susceptibility peak $\chi^p$ with increasing system size for pinned systems with $c_p\ge0.03$ indicates the absence of such first-order discontinuous transition.

Next, we investigate whether or not this brittle-to-ductile transition is separated by a critical point. For that, we choose the same order parameter as in Ref.~\cite{DuctileToBrittlePNAS}: the largest drop in stress $\Delta \sigma_{max}$ within the examined strain range. In the thermodynamic limit, $N\rightarrow \infty$, this quantity should vanish for ductile failure while remaining finite for brittle failure.
In Fig.~\ref{fig2}(a), the largest stress drop averaged over many realizations, $\langle \Delta \sigma_{max}\rangle$, is shown as a function of $c_p$ for different system sizes. For a fixed $N$, $\langle \Delta \sigma_{max}\rangle$ always decreases with increasing pinning concentration, owing to the enhanced ductility of pinned systems. However, the system size dependence of $\langle \Delta \sigma_{max}\rangle$, shown in Fig.~\ref{fig2}(b), changes with varying $c_p$. For unpinned and smaller $c_p$, we observe $\langle \Delta \sigma_{max}\rangle$ increases  and eventually saturates to a finite value whereas for larger $c_p$, $\langle \Delta \sigma_{max}\rangle$ decreases with system size. We observe a power law decay $\langle \Delta \sigma_{max}\rangle \sim N^{-0.5}$ for $c_p\ge0.01$, which supports the ductile failure in the thermodynamic limit. The crossover occurs roughly at the pinning concentration where $\langle \Delta \sigma_{max}\rangle$ remains insensitive to $N$. Interestingly, for very large system sizes $(N > 100000)$, $\langle \Delta \sigma_{max} \rangle$ decreases with $N$ for all $c_p$. This is due to the two-step relaxation of stress before reaching the steady state (see Fig.~\ref{fig1}). It does not affect the brittleness of the transition, as $\chi^p$ continues to increase.
%

To quantitatively demonstrate the critical nature of the transition, we study the system size dependence of the fluctuations in the maximum stress drop, $\chi_d = NVar(\Delta \sigma_{max}) = N( \langle \Delta \sigma_{max}^2\rangle - \langle \Delta \sigma_{max} \rangle^2)$.
In Fig.~\ref{fig2}(c), $\chi_d$ vs. $c_p$ curves are shown for different system sizes, which exhibit a non-monotonic variation with a sharp peak at an intermediate value of $c_p$. The peak value, $\chi_d^p$, is estimated by fitting $\chi_p$ vs $c_p$ curve with a Lorentzian function. With increasing $N$, $\chi_d^p$ increases, and the response becomes sharper, indicating the critical nature of the transition at a specific system-size-dependent critical value  $c_p^*(N)$.  The growth of $\chi_d^p$ is shown by the dashed line which diverge at $c_p = c_p^{*\infty}$ in the thermodynamic limit, where $\chi_d^p \rightarrow \infty$. In the inset $\langle \Delta \sigma_{max} \rangle$ is plotted using rescaled $c_p$ for better clarity of its rapid growth at higher $N$.
%
\begin{figure}
	\includegraphics[width=0.45\textwidth]{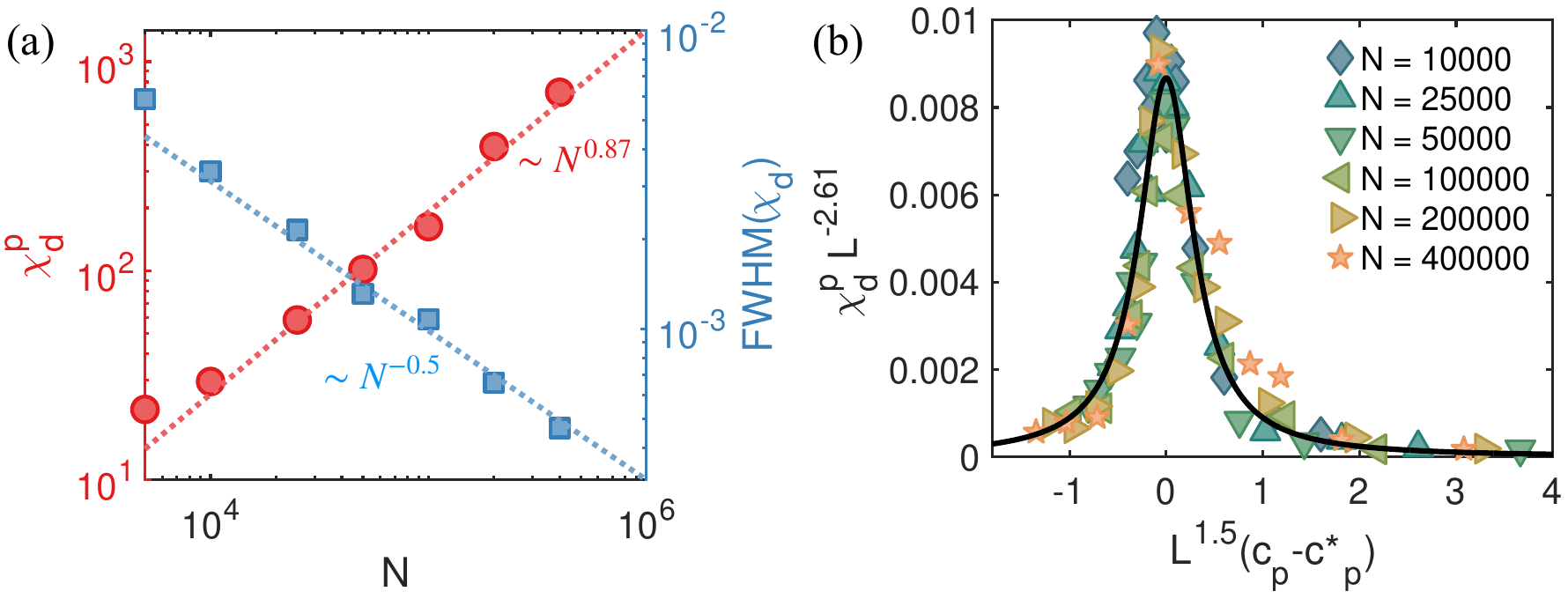}
	\caption{{\bf Critical exponents and data collapse.}
(a) Growth of the peak value of susceptibility $\chi_d^p$ (circle), and decay of FWHM (square) as a function of system size $N$. The red and blue dotted lines represent the best power law fit with exponents ${0.87\pm0.03}$ and ${-0.50\pm0.02}$, respectively. (b) Data collapse of $\chi_d$ vs. $c_p$ curves for different $N$ using obtained exponents (see text for details).  
	} 
	\label{fig3}
\end{figure}

Although the growth of the $\chi_d^p$ suggests the existence of a critical point at a critical $c_p^*$, it is necessary to perform finite-size scaling analysis to confirm it. Fig.~\ref{fig3}(a) shows the finite size scaling of $\chi_d^p$ and full width at half maxima (FWHM) of $\chi_d$ vs $c_p$ curves. The former exhibits a power law divergence in the thermodynamic limit following $\chi_d^p \sim N^{0.87\pm0.03}$, whereas the latter vanishes following $FWHM\sim N^{-0.50\pm0.02}$. This underpins the critical nature of brittle to ductile transition in our setup. We further explore this following the similar scaling ansatz used in Ref.~\cite{RossiRandomField} for AQS driven RFIM given by $\chi_d^p(c_p, L) \sim L^{\bar{\gamma}/\nu}\mathcal{F}(\bm{c}L^{1/\nu})$. Here $\bm c$ is the rescaled pinning concentration given by $\bm c = (c_p - c_p^{*})$. $\mathcal{F}(\bm{c}L^{1/\nu})$ is the scaling function with $\bm{c}L^{1/\nu}$ being the scaling variable. $\nu$, and $\bar{\gamma}$ are the critical exponents. This scaling form controls the height and spread of the $\chi_d$ vs. $c_p$ curves. According to the scaling ansatz, the peak height diverges as a power law of $L^{{\bar{\gamma}/\nu}}$, and FWHM vanishes with the power law of $ L^{-1/\nu}$. Given, exponents in Fig.~\ref{fig3}(a), we obtain, $\nu=0.67 \pm 0.03$, and $\bar{\gamma}/\nu=2.61 \pm 0.09$.  Exponent $\bar{\gamma}/\nu$ found in our simulations is very close to the one reported in Ref.~\cite{RossiEPMDToB} but $\nu$ is quite different for us. The data collapse using these exponents is shown in Fig.~\ref{fig3}(b). The quality of the collapse supports the chosen form of the scaling relation. 

%
The major criticism of the work in Ref.~\cite{RossiEPMDToB} is the existence of ductile failure in the thermodynamic limit. In Ref.~\cite{Fielding2020DtoB}, it has been claimed that, in the athermal quasi-static scenario, in the thermodynamic limit, all failures are brittle, suggesting the non-existence of a finite critical disorder. This question of having a ductile failure in the thermodynamic limit is also relevant to our setup. Here, without delving into the details of whether or not such a finite disorder critical point exists for samples prepared with different annealing protocols, we demonstrate that ductile failure indeed exists in the thermodynamic limit for highly disordered samples obtained by increasing the concentration of pinned particles with the following arguments. 
Firstly, the stress vs. strain curve for $c_p = 0.01$, shown in Fig.~\ref{fig1} (b), does not exhibit any system dependence. Secondly, in Fig.~\ref{fig2} (c), we do not observe any shift in the position of the maximum of $\chi_d$ vs. $c_p$ curve toward higher pinning concentrations. This suggests no possibility of $c_p^{*} \rightarrow \infty$ in the thermodynamic limit, eradicating the ductile phase. 
\begin{figure}
	\includegraphics[width=0.45\textwidth]{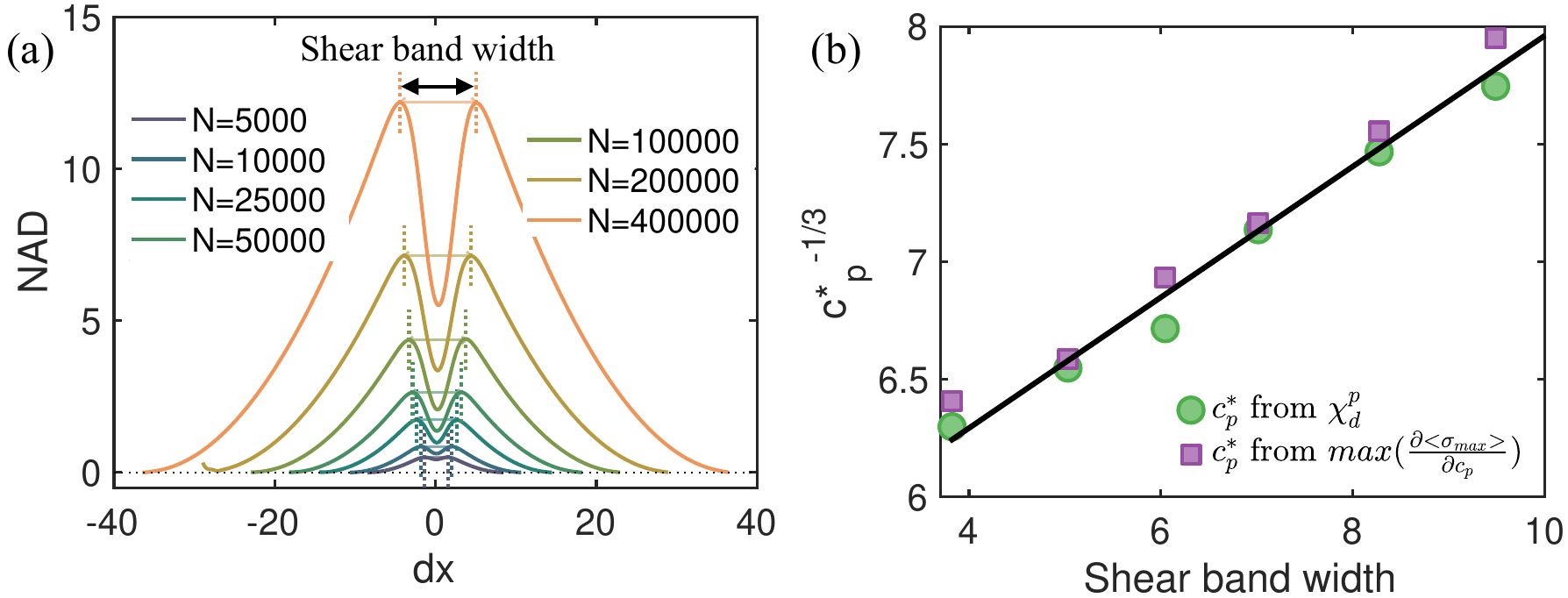}
	\caption{{\bf Link of $c_p^*$ to shear band width:}
		(a)  Non-affine displacement per particle (NAD), averaged over particles (and ensembles) in planes parallel to the shear band at a perpendicular distance $dx$ for different system sizes. The shear band region is identified by a central dip, giving the shear band width. (b)The growth of the shear band width with $N$ is compared to the growth of critical pinning length scale  $\xi_p \sim (c_p^*)^{-1/3}$, obtained from the peak of $\chi_d$, and $\partial(\Delta \sigma_{max})/\partial c_p$ .
	} 
	\label{fig4}
\end{figure}

However, interestingly, we observe a systematic shift of $c_p^*$ towards smaller $c_p$ (dashed line in Fig.~\ref{fig2} (c)), until the peak of the $\chi_d$  diverges at $c_p^{*\infty}$ (dotted line in Fig.~\ref{fig2} (c)). The point of divergence provides the value of critical pinning concentration in the thermodynamic limit.
This variation stems from the two competing length scales: the shear band width in the unpinned system and the pinning length scale, $\xi_p \sim (c_p^*)^{-1/3}$, suppressing that shear band. The shear band width increases with increasing system size (see Fig.~\ref{fig4} (a) )~\cite{ShangPengPinningPRB}, and should saturates to a finite value in the thermodynamic limit~\cite{PengPRE2022,WANG2016287}, explaining the reduction , and convergence of $c_p^*$ to $c_p^{*\infty}$. On the other hand, if the shear band width continues to increase with increasing system size then the critical disorder concentration will go to zero. In reality, shear bands have a finite thickness as reported in numerous experiments, one expects the shear band to be of finite width rather than infinite, indicating a strong argument for the existence of a finite disorder critical point in the yielding transition in amorphous solids. The direct correlation between these length scales is presented in Fig.~\ref{fig4} (b). 
%


\begin{figure}[!t]
	\includegraphics[width=0.4\textwidth]{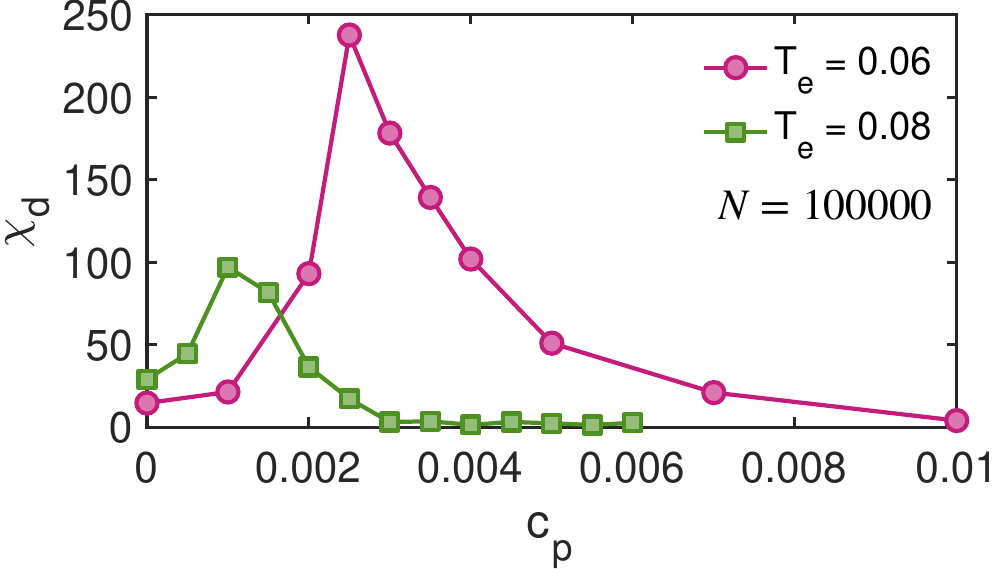}
	\caption{The critical pinning concentration $c_p^*$ shifts to smaller value for larger equilibration temperature.
	} 
	\label{fig5}
\end{figure}

At this point, if we concur with Ref.~\cite{RossiEPMDToB} that for a particular material,  transitions from ductile to brittle yielding occurs at a critical disorder strength of the samples, we can infer the following from our findings. Our samples, created by equilibrating the system at a temperature $T_e = 0.06$, are less disordered and undergo a brittle failure. By incorporating randomly pinned particles and adjusting their concentrations, we manipulate the disorder strength of the samples. At $c_p = c_p^*$, the samples attain the required critical disorder strength, while for higher concentrations, such as $c_p = 0.1$, they are significantly disordered and show ductile yielding. 
%
This implies that depending on the degree of disorder in our freshly prepared samples, the required critical disorder can be attained by an appropriate $c_p$ value. To verify this, we equilibrate samples at $T_e = 0.08$, which are brittle but have higher intrinsic disorder strength compared to those at $T_e = 0.06$. Consequently, the transition is expected to occur at lower $c_p$ values. The results from these simulations are presented in Fig.~\ref{fig5}, which shows that $\chi_d$ peaks at smaller $c_p^{*}$ for higher $T_e$, further supporting the reliability of our inference.
\vskip +0.1in
\noindent{\bf Conclusions:}
To conclude, we study the brittle-to-ductile transition in yielding of amorphous solids with increasing quenched disorder introduced in the form of random particle pinning. Two states are identified using an order parameter, the largest plastic stress drop in the thermodynamic limit, which vanishes in the ductile state but remains finite in the brittle state. Our study reveals that this transition is a non-equilibrium phase transition, with sample to sample fluctuations in the order parameter diverging at a critical pinning concentration, corresponding to a critical disorder strength. We demonstrate that by controlled adding pinned particles, we increase the degree of disorder of the sample systematically, and the transition occurs when the degree of disorder reaches the critical value. The critical exponents obtained in our study are close to those found for the athermal, quasi-statically driven Random Field Ising Models with Eshelby interactions. However, whether they belong to the same universality class requires further exploration. It is important to note that our findings are limited to conditions of vanishing temperature and shear rate. The introduction of a finite shear rate or temperature would likely to alter the results~\cite{MurariBToD, SurajitPNAS,LemaitreAndCaroli}. 

\vskip +0.1in
\noindent{\bf Acknowledgements:}
We would like to thank Gilles Tarjus and Christopher Ness for many useful discussions. We acknowledge funding by intramural funds at TIFR Hyderabad from the Department of Atomic Energy (DAE) under Project Identification No. RTI 4007. SK acknowledges the Swarna Jayanti Fellowship grants DST/SJF/PSA01/2018-19 and SB/SFJ/2019-20/05 from the Science and Engineering Research Board (SERB) and Department of Science and Technology (DST). SK also acknowledges research support from MATRICES Grant MTR/2023/000079 from SERB.
\bibliography{ALL}
\end{document}